\documentclass{emulateapj}
\slugcomment{} 
\shorttitle{Third image of SDSS J1029+2623}
\shortauthors{Oguri et al.}
\begin{document}
%%%%%%%%%%%%%%%%%%%%%%%%%%%%%%%%%%%%%%%%%%%%%%%%%%%%%%%%%%%%%%%%%%%%%%
%%%%%%%%%%%%%%%%%%%%%%%%%%%%%%%%%%%%%%%%%%%%%%%%%%%%%%%%%%%%%%%%%%%%%%
\title{The Third Image of the Large-Separation Lensed Quasar 
SDSS~J1029+2623\altaffilmark{1}}   
%%%%%%%%%%%%%%%%%%%%%%%%%%%%%%%%%%%%%%%%%%%%%%%%%%%%%%%%%%%%%%%%%%%%%%
%%%%%%%%%%%%%%%%%%%%%%%%%%%%%%%%%%%%%%%%%%%%%%%%%%%%%%%%%%%%%%%%%%%%%%
%
%%%%%%%%%%%%%%%%%%%%%%%%%%%%%%%%%%%%%%%%%%%%%%%%%%%%%%%%%%%%%%%%%%%%%%
\author{
Masamune Oguri,\altaffilmark{2} 
Eran O. Ofek,\altaffilmark{3} 
Naohisa Inada,\altaffilmark{4} 
Tomoki Morokuma,\altaffilmark{5} 
Emilio E. Falco,\altaffilmark{6} \\
Christopher S. Kochanek,\altaffilmark{7}
Issha Kayo,\altaffilmark{8}
Tom Broadhurst,\altaffilmark{9} and
Gordon T. Richards\altaffilmark{10}
}

\altaffiltext{1}{Some of the data presented herein were obtained
at the W.M. Keck Observatory, which is operated as a scientific
partnership among the California Institute of Technology, the University
of California and the National Aeronautics and Space Administration. The
Observatory was made possible by the generous financial support of the
W.M. Keck Foundation. This work is based in part on data collected at
Subaru Telescope, which is operated by the National Astronomical
  Observatory of Japan.  Use of the UH 2.2-m telescope for the
  observations is supported by NAOJ.}

\altaffiltext{2}{Kavli Institute for Particle Astrophysics and
 Cosmology, Stanford University, 2575 Sand Hill Road, Menlo Park, CA 
 94025.}  
\altaffiltext{3}{Division of Physics, Mathematics and Astronomy,
 California Institute of Technology, Pasadena, CA 91125.}
\altaffiltext{4}{Cosmic Radiation Laboratory, RIKEN (The Institute of
 Physical and Chemical Research), 2-1 Hirosawa, Wako, Saitama
 351-0198, Japan.}  
\altaffiltext{5}{Optical and Infrared Astronomy Division, National
 Astronomical Observatory of Japan, Mitaka, Tokyo, 181-8588, Japan.}
\altaffiltext{6}{Harvard-Smithsonian Center for Astrophysics,
 Cambridge, MA 02138.} 
\altaffiltext{7}{Department of Astronomy, Ohio State University,
 Columbus, OH 43210.} 
\altaffiltext{8}{Department of Physics and Astrophysics, Nagoya
University, Chikusa-ku, Nagoya 464-8602, Japan.} 
\altaffiltext{9}{School of Physics and Astronomy, Tel Aviv University,
 Tel Aviv 69978, Israel.} 
\altaffiltext{10}{Department of Physics, Drexel University, 3141
                 Chestnut Street,  Philadelphia, PA 19104.}

\setcounter{footnote}{10}

\begin{abstract}
We identify a third image in the unique quasar lens SDSS J1029+2623, 
the second known quasar lens produced by a massive cluster of
galaxies. The spectrum of the third image shows similar emission and
absorption features, but has a redder continuum than the other two
images which can be explained by differential extinction or
microlensing. We also identify several lensed arcs. Our observations
suggest a complicated structure of the lens cluster at $z\approx 0.6$.
We argue that the three lensed images are produced by a naked cusp on
the basis of successful mass models, the distribution of cluster
member galaxies, and the shapes and locations of the lensed arcs. 
Lensing by a naked cusp is quite rare among galaxy-scale lenses but is
predicted to be common among large-separation lensed quasars. Thus the
discovery can be viewed as support for an important theoretical
prediction of the standard cold dark matter model. 
\end{abstract}

\keywords{galaxies: clusters: general --- 
gravitational lensing --- quasars: individual
(SDSS~J102913.94+262317.9)}  

%%%%%%%%%%%%%%%%%%%%%%%%%%%%%%%%%%%%%%%%%%%%%%%%
%%%%%%%%%%%%%%%%%%%%%%%%%%%%%%%%%%%%%%%%%%%%%%%%
%%%%%%%%%%%%%%%%%%%%%%%%%%%%%%%%%%%%%%%%%%%%%%%%
\section{Introduction}
%%%%%%%%%%%%%%%%%%%%%%%%%%%%%%%%%%%%%%%%%%%%%%%%
%%%%%%%%%%%%%%%%%%%%%%%%%%%%%%%%%%%%%%%%%%%%%%%%
%%%%%%%%%%%%%%%%%%%%%%%%%%%%%%%%%%%%%%%%%%%%%%%%

SDSS~J102913.94+262317.9  \citep[SDSS J1029+2623;][]{inada06} is one of
two examples of strongly lensed quasars produced by massive clusters of
galaxies. It was discovered in the Sloan Digital Sky Survey Quasar Lens
Search \citep[SQLS;][]{oguri06,oguri08,inada08}, a survey to identify
gravitationally lensed quasars from the spectroscopic sample of 
quasars in the Sloan Digital Sky Survey \citep[SDSS;][]{york00}. The
image separation of $22\farcs5$ makes it the largest lensed quasar
known to date.  \citet{inada06} found that the system consists of two
images of a radio-loud quasar at $z=2.197$ created by a massive cluster
of galaxies at $z\sim 0.6$.   The rareness of such quasar-cluster lens
systems \citep[e.g.,][]{ofek01,phillips01,inada03} validates the importance of
understanding this system with extensive follow-up work.  

In this {\it Letter}, we present new imaging and spectroscopic
observations of this lens system. Based on these observations we
identify a third lensed quasar image, measure the redshift of the lens
cluster, and identify several additional lensed arcs. We discuss the
observations in \S~\ref{sec:obs}, model and interpret the lens system 
in \S~\ref{sec:model}, and summarize our results in \S~\ref{sec:sum}.

%%%%%%%%%%%%%%%%%%%%%%%%%%%%%%%%%%%%%%%%%%%%%%%%
%%%%%%%%%%%%%%%%%%%%%%%%%%%%%%%%%%%%%%%%%%%%%%%%
%%%%%%%%%%%%%%%%%%%%%%%%%%%%%%%%%%%%%%%%%%%%%%%%
\section{Follow-up observations}\label{sec:obs}
%%%%%%%%%%%%%%%%%%%%%%%%%%%%%%%%%%%%%%%%%%%%%%%%
%%%%%%%%%%%%%%%%%%%%%%%%%%%%%%%%%%%%%%%%%%%%%%%%
%%%%%%%%%%%%%%%%%%%%%%%%%%%%%%%%%%%%%%%%%%%%%%%%

We obtained deep images of the system using the University of Hawaii
2.2-meter (UH88) and Keck I telescopes. In the UH88 observations we took
1000~s B-band, 1200~s $VRI$-band, and 400~s $VRI$-band images on the
nights of 2006 November 14, 2007 May 17, and 2007 November 12 using the 
Tektronix 2048$\times$2048 CCD camera (Tek2k). The typical
seeing was $\sim 0 \farcs8$ and the conditions were photometric.
The photometry was calibrated by the standard star PG 0918+029
\citep{landolt92}. We obtained additional 720~s $g$-band and 880~s
$R$-band images with the Keck I telescope using the Low Resolution
Imaging Spectrometer with the Atmospheric Dispersion Compensator
\citep[LRIS-ADC;][]{oke95} on 2008 January 4 when the seeing was 
$\sim 0\farcs7$. Since no standard star was observed, we calibrated the
images using the UH88 and SDSS images. These data were reduced and
analyzed using standard IRAF tasks.  

%%%%%%%%%%%%%%%%%%%%%%%%%%%%%%%%%%%%%%%%%%%%%%%%%%%%%%%%%%%%%%%%%%%%%%%
\begin{figure*}
\epsscale{.9}
\plotone{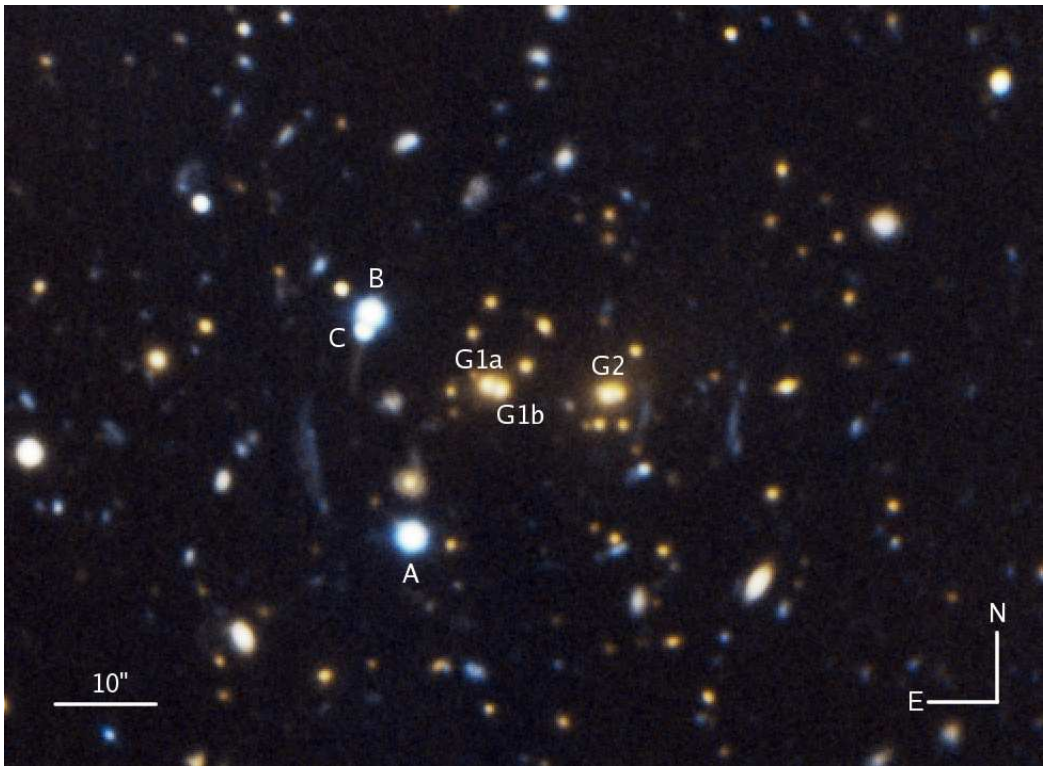}
\caption{Follow-up color image of SDSS J1029+2623 produced from the
 Keck $g$- and $R$-band images. The positions, magnitudes, and redshifts
 of the lensed images and cluster members are summarized in 
 Table~\ref{tab:pos}.  Several lensed arcs are also seen. Note
 particularly two large blue arcs on the east and west sides of galaxies
 G1/G2, a red arc on the south of object C, and a blue arc located near
 galaxy G2.   
\label{fig:img1029}}
\end{figure*}
%%%%%%%%%%%%%%%%%%%%%%%%%%%%%%%%%%%%%%%%%%%%%%%%%%%%%%%%%%%%%%%%%%%%%%%

%%%%%%%%%%%%%%%%%%%%%%%%%%%%%%%%%%%%%%%%%%%%%%%%%%%%%%%%%%%%%%%%%%%%%%
\begin{deluxetable*}{crrccccccc}
\tablewidth{0pt}
%\rotate
\tablecaption{SDSS J1029+2623: Astrometry and photometry}
\tablewidth{0pt}
\tablehead{\colhead{Name} & 
 \colhead{$\Delta x$ [${}''$]} &
 \colhead{$\Delta y$ [${}''$]} &
 \colhead{$B$(UH)} & \colhead{$V$(UH)} & \colhead{$R$(UH)} &
 \colhead{$I$(UH)} & \colhead{$g$(Keck)} & \colhead{$R$(Keck)} &
 \colhead{Redshift} }
\startdata
A   & $0.00$  & $0.00$  & 19.20   & 18.72 & 18.55 & 18.01 & 18.72 & 18.46 & 2.197 \\
B   & $-3.87$ & $22.19$ & 19.03   & 18.67 & 18.51 & 17.95 & 18.81 & 18.58 & 2.197 \\
C   & $-4.53$ & $20.55$ & 20.89   & 20.63 & 20.28 & 19.51 & 20.87 & 20.38 & 2.197 \\
G1a & $7.89$  & $14.90$ & $>$23.5 & 22.18 & 20.92 & 19.64 & 22.98 & 20.93 & 0.596 \\
G1b & $8.93$  & $14.54$ & $>$23.5 & 22.34 & 21.12 & 19.73 & 23.01 & 21.02 & 0.601 \\
G2  & $19.85$ & $14.00$ & $>$23.5 & 22.00 & 20.62 & 19.26 & 22.69 & 20.60 & 0.584 \\
\enddata
\tablecomments{The relative positions are from the UH88 $I$-band image. 
 Note that the Equatorial J2000.0 coordinates of object A are 10:29:13.94
 +26:23:17.9. The UH88 images were taken on 2006  November ($B$) and
 2007 May ($VRI$), whereas the Keck images were obtained on 2008 January 4. 
 The magnitudes are  $3''$ diameter aperture magnitudes. Measurement
 errors on the positions and magnitudes are typically $\lesssim
 0\farcs01$ and $\lesssim 0.02$~mag, respectively. All the redshifts are
 spectroscopic, from \citet{inada06} (A and B), Keck (B and C) and
 Subaru (G1a, G1b, G2).}    
\label{tab:pos}
\end{deluxetable*}
%%%%%%%%%%%%%%%%%%%%%%%%%%%%%%%%%%%%%%%%%%%%%%%%%%%%%%%%%%%%%%%%%%%%%%%

We show the Keck image in Figure \ref{fig:img1029}. The results of the
photometry and astrometry are summarized in Table \ref{tab:pos}. 
Several lensed arcs are clearly visible in the image. As suggested in 
\citet{inada06}, the central galaxy G1 consists of two components,
which we name G1a and G1b. \citet{inada06} argued that  object C is not
a lensed image on the basis of its different color from image A and B,
and our follow-up results confirm that the color of object C is
significantly different from those of the other lensed quasar images. 
However, both by comparing the UH88 $V$-band images and from an ongoing 
monitoring at the Fred Lawrence Whipple Observatory 1.2-meter telescope
\citep[see][]{fohlmeister07,fohlmeister08}, we see that both B and C
have faded by approximately 0.1~mag during 2007. This suggests that
object C is a quasar, as variable blue sources at this magnitude level
are typically quasars \citep[e.g.,][]{sesar07}. 

We obtained spectra of objects B and C with the LRIS-ADC at the Keck I
telescope on 2007 December 14, and spectra of galaxies G1a, G1b, and G2
with the Faint Object Camera and Spectrograph
\citep[FOCAS;][]{kashikawa02} at the Subaru telescope on 2007 January
23. In the Keck LRIS-ADC observation, the 800~s exposure was taken with
a long slit ($1\farcs0$ width) aligned with objects B and C. We used the
D560 dichroic with the 400/8500 grating and the 400/3500 grism to achieve
a spectral resolution of $\sim 5$\AA. The observations were obtained
in non-photometric conditions with poor $\sim 2\farcs0$ seeing. We used
the deblending procedure of \citep{pindor06} in which we assumed Gaussian
profiles for spatial cross sections  to extract independent spectra
of the two components. The standard star Feige 34 \citep{oke90} was used
for flux calibration. The Subaru FOCAS observation was conducted in
$2\times 2$ on-chip binning mode with the 300B grism and the SY47
filter.  The $900$~s exposure was taken in $\sim 0\farcs6$ seeing, with
a long slit ($1\farcs0$ width) aligned across the galaxies G1a, G1b, and
G2. The spectral resolution with this setup is $R\sim 400$.

\begin{figure}
\epsscale{.99}
\plotone{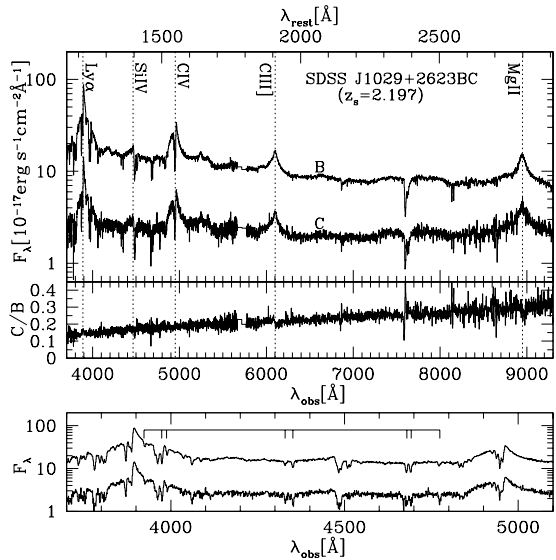}
\caption{{\it Upper}: Spectra of components B and C obtained with the
  LRIS-ADC at Keck. Quasar emission lines redshifted to $z_s=2.197$
  are indicated by vertical dotted lines. The ratio of the spectra is
  also shown.
{\it Lower}: Expanded view of the blue side to show absorption systems 
 more clearly. In addition to strong absorption lines associated
with the emission lines, a \ion{Mg}{1}/\ion{Mg}{2}/\ion{Fe}{2} absorption
 system at $z=0.674$, which is indicated by solid bars, is seen in
  both B and C. We note that additional \ion{Mg}{2} absorptions at
  $z=1.910$ (in B and C) and $z=1.761$ (in B) are also detected in the
  spectra. 
\label{fig:spec1029}}
\end{figure}

Figure~\ref{fig:spec1029} unambiguously shows that component C is a
quasar with the same redshift, $z_s=2.197$, as component B. The shapes
of the emission and absorption lines are similar. Note in
particular similar absorptions in the wings of the Ly$\alpha$,
\ion{Si}{4}, and \ion{C}{4} lines. These absorption features
are probably intrinsic to the quasar and therefore serve as strong
evidence that objects B and C are lensed images of the same quasar. As
we expect from the different broad band colors, image C has a redder
continuum than image B. A possible explanation is differential
extinction, which is relatively common in lensed quasars
\citep[e.g.,][]{falco99,eliasdottir06}. This requires that the two
paths have  $\Delta E(B-V)\sim 0.15-0.2$ assuming the extinction is
caused by dust ($R_V=3.1$) at the cluster redshift, $z\approx 0.6$. 
We subtracted models for images B and C from the Keck images and found
no nearby galaxies beyond the red arc to the south of image C.
Another possibility is microlensing by stars in the lens, which can
also modify the color of the continuum
\citep[e.g.,][]{poindexter07,ofek07}. We note that both dust reddening
and microlensing are observed in the other cluster-scale lensed quasar
SDSS~J1004+4112 \citep{oguri04b,richards04,fohlmeister07,fohlmeister08}.  
Moreover, the Hubble Space Telescope image of this system also shows
no obvious candidate galaxy for the effects. In principle we can
distinguish between  these possibilities by examining the differences
in the flux ratios between the continuum and the emission lines,
because dust will affect both equally while microlensing affects the
compact continuum emission region more strongly than the larger broad
line regions \citep[e.g.,][]{keeton06}. The ratio of the spectra shown
in Figure~\ref{fig:spec1029} shows no strong difference between
continuum and emission line regions. However, we note a possible small
difference in the \ion{Mg}{2} emission line which suggests that
microlensing is affecting this system.   

%%%%%%%%%%%%%%%%%%%%%%%%%%%%%%%%%%%%%%%%%%%%%%%%%%%%%%%%%%%%%%%%%%%%%%%
%%%%%%%%%%%%%%%%%%%%%%%%%%%%%%%%%%%%%%%%%%%%%%%%%%%%%%%%%%%%%%%%%%%%%%%

Finally, the redshifts of G1a, G1b, and G2 are $z=0.596$,
$0.601$, and $0.584$, respectively, confirming the estimated cluster
redshift of $z \approx 0.6$. However, the velocity difference between 
G1b and G2 of $2800$~${\rm km\,s^{-1}}$ is so large that the system
may be an on-going merger or a superposition of two smaller clusters. 

%%%%%%%%%%%%%%%%%%%%%%%%%%%%%%%%%%%%%%%%%%%%%%%%
%%%%%%%%%%%%%%%%%%%%%%%%%%%%%%%%%%%%%%%%%%%%%%%%
%%%%%%%%%%%%%%%%%%%%%%%%%%%%%%%%%%%%%%%%%%%%%%%%
\section{Interpreting the lens system}\label{sec:model}
%%%%%%%%%%%%%%%%%%%%%%%%%%%%%%%%%%%%%%%%%%%%%%%%
%%%%%%%%%%%%%%%%%%%%%%%%%%%%%%%%%%%%%%%%%%%%%%%%
%%%%%%%%%%%%%%%%%%%%%%%%%%%%%%%%%%%%%%%%%%%%%%%%

Our identification of the third image has several important
implications for the mass distribution of the lensing
cluster. There are four possible explanations for image C: (i) it
is a so-called central core image \citep{winn04,inada05} of a
``double'' lens system; (ii) it is a ``double'' lens system, but one
of the images is split into a pair of images because of external
perturbations \citep[e.g.,][]{schechter02,keeton03}; (iii) the lens
system is a  standard cusp-type ``quadruple'' lens, but the fourth
(counter) image is too faint to be observed; and (iv) the three images
are produced by a naked cusp.

The main problem for (i) and (ii) is that the models require a large
offset between the positions of the brightest cluster galaxies and the
center of the lens potential, as discussed in \citet{inada06}. In
addition, the lensed arcs seem to contradict these scenarios as their
shapes and locations suggest the lens potential is centered near G1
and G2 rather than somewhere between A and B.  To test (iii), we
searched for any blue objects defined by $B-V<0.7$, $V-R<0.6$, and
$R-I<1.0$ in the UH88 images. We expect the color cut is conservative
as it includes image C which is significantly reddened. Except the
three confirmed lensed images, we found no such blue object, to the
limiting magnitude of $B\sim 23.5$, within $60''$ of image A. This
translates into a constraint on the flux ratio of the putative fourth
image D of $D/A\lesssim 0.02$, which is strong enough to make scenario
(iii) unlikely. We also searched for additional variable sources in
the monitoring data and found none near the cluster, although with the
present data we would not detect variability in a source significantly
fainter than the known images.    

%%%%%%%%%%%%%%%%%%%%%%%%%%%%%%%%%%%%%%%%%%%%%%%%%%%%%%%%%%%%%%%%%%%%%%%
\begin{figure}
\epsscale{.9}
\plotone{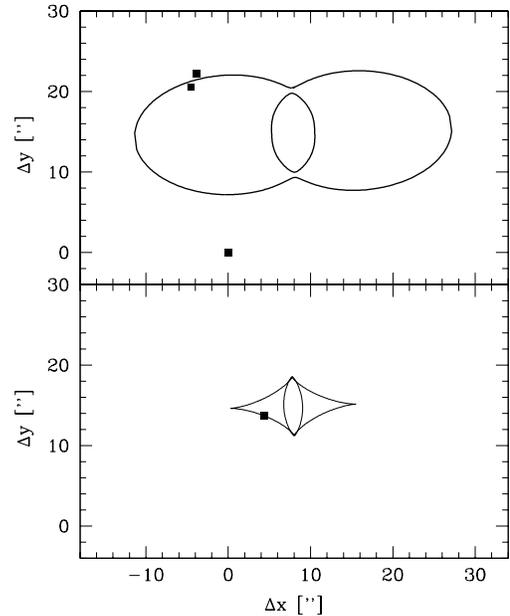}
\caption{Critical curves ({\it upper}) and caustics ({\it lower}) of the
  best-fit model reproducing the three image positions. The square in
 the lower panel shows the best-fit source position, whereas the three
 squares in the upper panel indicate the corresponding best-fit image
 positions which are very close to the observed image positions. Because
 of the naked cusp in the caustics, this model predicts only three
 images on the same side of the lens potential, which explains the
 unique image configuration of this lens system (see also Figure
 \ref{fig:img1029}). The model is an elliptical \citet{navarro97}
 profile with virial mass $M_{\rm  vir}= 1.2\times 10^{15}M_\odot$,
 concentration parameter $c_{\rm vir}= 4.9$, ellipticity $e=0.44$, and
 position angle $\theta_e=-88^\circ$. However, note that the derived
 mass and concentration parameter crucially depend on our assumption of
 the scale radius, $r_s=60''$.
\label{fig:cusp}}
\end{figure}
%%%%%%%%%%%%%%%%%%%%%%%%%%%%%%%%%%%%%%%%%%%%%%%%%%%%%%%%%%%%%%%%%%%%%%%

The remaining scenario, (iv), appears to be a viable solution. A naked
cusp is a rare configuration of caustics in which a tangential caustic
extends outside a radial caustic. In the case that a source is inside a
naked cusp, three bright images are created on the same side of the
potential center. For definiteness, we fit the positions of the three
images using the {\it lensmodel} package \citep{keeton01}. We adopt an
elliptical \citet{navarro97} density profile centered on galaxy G1a with
a fixed scale of $60''$. We note that the scale radius is degenerate
with the estimated mass \citep{oguri04b}. We assume a positional error
of $0\farcs5$ to allow for perturbations from substructures (e.g.,
galaxies) in the cluster. This simple model, shown in
Figure~\ref{fig:cusp}, works surprisingly well, with $\chi^2=8\times
10^{-3}$ for 1 degree of freedom.  The position angle of the model
($-88^\circ$ East of North) is reasonable since it is aligned with
the G1-G2 axis and the spatial distribution of color-selected cluster
members, and puts the lensed arcs on the major axis of the lens
potential as is preferred by ray-tracing in $N$-body simulations
\citep{dalal04}. This model predicts time delays between the AB and BC
images of $1860h^{-1}$~days and $2.3h^{-1}$~days, respectively,
assuming a cosmological model of $\Omega_M=0.3$ and
$\Omega_\Lambda=0.7$.   

A caveat for this model, as in many models of gravitational
lenses, is the flux ratios of the quasar images. Given the image
geometry, we would generically expect $B \approx C > A$, while we
observe $B \approx A >C$. Specifically, our model predicts
$A:B:C=0.11:1.00:0.99$, whereas the observed flux ratios are
$A:B:C=0.95:1.00:0.24$ in the UH88 $I$-band image. The strong wavelength
dependence of the flux ratio between images B and C, which is probably
because of dust and/or microlensing, needs to be understood in order to
fully evaluate this problem.  We also note that the A/B flux ratio in
the 6-cm Very Large Array (VLA) radio map of $\sim 0.73$ \citep{inada06}
is quite different from the ratio in the optical, $\sim 0.95$. The
spatial resolution and signal-to-noise ratio of the radio map were not
sufficient to derive robust radio fluxes for all the three components,
and thus additional deep, high-resolution radio images will be essential
for understanding the anomalous flux ratios.   

%%%%%%%%%%%%%%%%%%%%%%%%%%%%%%%%%%%%%%%%%%%%%%%%
%%%%%%%%%%%%%%%%%%%%%%%%%%%%%%%%%%%%%%%%%%%%%%%%
%%%%%%%%%%%%%%%%%%%%%%%%%%%%%%%%%%%%%%%%%%%%%%%%
\section{Summary}\label{sec:sum}
%%%%%%%%%%%%%%%%%%%%%%%%%%%%%%%%%%%%%%%%%%%%%%%%
%%%%%%%%%%%%%%%%%%%%%%%%%%%%%%%%%%%%%%%%%%%%%%%%
%%%%%%%%%%%%%%%%%%%%%%%%%%%%%%%%%%%%%%%%%%%%%%%%

In \citet{inada06} we were left with a puzzle about SDSS~J1029+2623: 
the observed geometry seemed inconsistent with the observed cluster
center.   Our discovery here that object C is a third image of the
quasar, despite its redder continuum, solves that problem because  
a three-image cusp lens configuration is easily produced under these
circumstances.  The next problem is to understand the flux ratios of
the system, both the wavelength dependences and that we observe $B
\approx A >C$ rather than the expect $B \approx C > A$. Measuring
the radio flux ratios will let us separate the effects of substructures
in the cluster from extinction and microlensing, and the
differences between continuum and emission line, radio flux ratios,
and time variability will allow us to separate the effects of
extinction and microlensing.

\lastpagefootnotes

SDSS J1029+2623 is probably the second naked-cusp quasar lens.
Among the $\sim 100$ galaxy-scale lenses, only one candidate is known
\citep[APM 08279+5255;][]{lewis02}. Naked cusps require ``marginal
lenses'', in which the lens has a surface density only moderately above
the critical surface density required for the production of multiple
images \citep[see][]{blandford87}. The cooling of the baryons and the
formation of the stellar component of a galaxy provides such high
central surface densities that it is nearly impossible for a galaxy to
have a naked cusp unless it is an edge-on disk-dominated system
\citep[see][]{keeton98}. Group and cluster halos, where the baryons have
not cooled yet, are far less efficient lenses 
\citep[e.g.,][]{kochanek01} and naked cusp configurations should be far
more common. Specifically, 
\citet[][see also \citealt{li07,minor08}]{oguri04a} predicted that
$30\%-60\%$ of cluster-scale lensed quasars will be naked-cusp systems
depending on the inner density profile.\footnote{While cusp geometries
also appear to be common for galaxies lensed by clusters, the more
complex selection functions and the difficulty of identifying all the
lensed images make it harder to do statistics.}  
We cannot seriously measure these fractions at present, given one
naked cusp lens and one conventional quadruple lens \citep[SDSS
  J1004+4112;][]{inada03}, but the existence of even one naked-cusp
cluster lens at this stage supports an important prediction of
standard cold dark matter halos.  

\acknowledgments

We thank Paul Schechter for useful discussions, and an anonymous
referee for helpful suggestions. 
This work was supported in part by Department of Energy contract
DE-AC02-76SF00515. CSK is supported by NSF grand AST-0708082.
The authors wish to recognize and acknowledge the very significant
cultural role and reverence that the summit of Mauna Kea has always
had within the indigenous Hawaiian community.  We are most fortunate
to have the opportunity to conduct observations from this mountain.

\end{document}